\documentclass[12pt]{article}

\usepackage{mathtools}
\usepackage{booktabs}
\usepackage[english]{babel} % English language/hyphenation
\usepackage[protrusion=true,expansion=true]{microtype} % Better typography
\usepackage{amsmath,amsfonts,amsthm}
\usepackage{ amssymb }
\usepackage{enumitem}
\usepackage{graphicx}
\usepackage{array}
\usepackage[toc,page]{appendix}
\usepackage{xcolor}
\usepackage{comment}
\usepackage{subfig}
\usepackage{algorithm}
\usepackage{algpseudocode}

\usepackage{tikz}
\usetikzlibrary{arrows.meta, positioning}
\usetikzlibrary{matrix, positioning}
\usepackage{booktabs} % Horizontal rules in tables
\usepackage{natbib}
\usepackage{setspace}
\usepackage{microtype} % Slightly tweak font spacing for aesthetics
\usepackage{tabularx}
\usepackage{subcaption}
\usepackage[font = small,labelfont=bf,textfont=it]{caption} % Custom captions under/above floats in tables or figures
\usepackage{footnote}
\usepackage{algorithm}% http://ctan.org/pkg/algorithms
\usepackage{algpseudocode}% http://ctan.org/pkg/algorithmicx
\usepackage{multirow}
\usepackage{hyperref}
\usepackage{cleveref}
\makeatletter
\newcommand{\distas}[1]{\mathbin{\overset{#1}{\kern\z@\sim}}}%
\newsavebox{\mybox}\newsavebox{\mysim}

\theoremstyle{definition}

\newcommand{\distras}[1]{%
  \savebox{\mybox}{\hbox{\kern3pt$\scriptstyle#1$\kern3pt}}%
  \savebox{\mysim}{\hbox{$\sim$}}%
  \mathbin{\overset{#1}{\kern\z@\resizebox{\wd\mybox}{\ht\mysim}{$\sim$}}}%
}
\newcolumntype{C}[1]{>{\centering\let\newline\\\arraybackslash\hspace{0pt}}m{#1}}

\newcommand{\vect}[1]{\ensuremath{\mathbf{#1}}}

\newcommand{\blind}{1}

\begin{document}

\def\spacingset#1{\renewcommand{\baselinestretch}%
{#1}\small\normalsize} \spacingset{1.3}

%%%%%%%%%%%%%%%%%%%%%%%%%%%%%%%%%%%%%%%%%%%%%%%%%%%%%%%%%%%%%%%%%%%%%%%%%%%%%%

\if1\blind
{
 \centering{\bf\Large Influence of Prior Distributions on Gaussian Process Hyperparameter Inference
 %GP priors in MCMC comparison (tentative title, you can change if you have an idea; The Role of Priors (and Proposals) in Bayesian Inference for Gaussian Process covariance parameter / Influence of Priors (and Proposals) on Gaussian Process covariance parameter Inference: A Bayesian Perspective / An Empirical Study on Prior Sensitivity in MCMC-Based Gaussian Process Modeling)
 }\\
  \vspace{0.3in}
  \centering{Ayumi Mutoh \vspace{0.1in}\\
        Department of Statistics, North Carolina State University\\
  and\\
  Junoh Heo\footnote{Corresponding author. \href{mailto:heojunoh@msu.edu}{heojunoh@msu.edu}} \vspace{0.1in}\\
  Department of Statistics and Probability, Michigan State University}
        
    \date{\vspace{-7ex}}
  %\maketitle
} \fi

\if0\blind
{
  \bigskip
  \bigskip
  \bigskip
    \begin{center}
    {\LARGE\bf }
\end{center}
  \medskip
} \fi

\bigskip
\begin{abstract}
Gaussian processes (GPs) are widely used metamodels for approximating expensive computer simulations, particularly in engineering design and spatial prediction. However, their performance can deteriorate significantly when covariance parameters are poorly estimated, highlighting the importance of accurate inference. The most common approach involves maximizing the marginal likelihood, yielding point estimates of these parameters. However, this approach is highly sensitive to initialization and optimization settings. An alternative is to adopt a fully Bayesian hierarchical framework, where the posterior distribution over the covariance parameters is inferred. This approach provides more robust uncertainty quantification and reduces sensitivity to parameter selection. Yet, a key challenge lies in the careful specification of prior distributions for these parameters. While many available software packages provide default priors, their influence on model behavior is often underexplored. Additionally, the choice of proposal distributions can also influence sampling efficiency and convergence. In this paper, we examine how different prior and proposal distributions over the lengthscale parameters $\theta$ affect predictive performance in a hierarchical GP model, using both simulated and real data experiments. By evaluating various types of priors and proposals, we aim to better understand their influence on predictive accuracy and uncertainty quantification. 
\end{abstract}

\noindent%
\vfill

\newpage
\spacingset{1.45} % DON'T change the spacing!

\section{Introduction}\label{sec:intro}
With the advancement of computer simulations, Gaussian processes (GPs) have proven to be powerful and effective tools for non-linear modeling. Their flexibility in capturing complex relationships in data, strong interpolation capability, and interpretability have helped them gain significant popularity across diverse fields, such as machine learning, spatial statistics, and computer experiments \citep{rasmussen2006gaussian, banerjee2014spatial, santner2018design, Harshvardhan2019compsim}. This popularity has led to the development of various software packages designed for GPs in R, Python, JMP, and MATLAB. 

The most widely used approach for inference in GPs is deterministic approximate inference \citep{rasmussen2006gaussian, Rue2009inference}. This method requires point estimates of the model's covariance parameters. A common strategy for obtaining these estimates is to maximize the marginal likelihood function, yielding the maximum likelihood estimates (MLEs). These MLEs are then treated as fixed values and substituted into subsequent computations. Several software packages adopt this plug-in approach as their default setting, including \texttt{mlegp} \citep{garrett2013mlegp} and \texttt{DiceKriging} \citep{roustant2012dicekriging} in R. However, this plug-in approach ignores uncertainty in the covariance parameters. This limitation becomes more problematic in complex GP settings, such as when data are sparse or noisy, where point estimates can lead to poor fits, instability, or underestimation of uncertainty. To address this issue, some researchers adopt Bayesian methods that explicitly model uncertainty in the covariance parameters. Among these, full Bayesian inference offers a more principled and rigorous alternative by accounting for this uncertainty \citep{Titsias2011mcmc}. 

In a fully Bayesian framework, the posterior distribution of the covariance parameters is inferred within a hierarchical model structure \citep{paulo2005bayes, lalchand2020bayes, riis2023bayes}. This is typically conducted using Markov chain Monte Carlo (MCMC), which places priors on the covariance parameters and draws samples from their posterior distribution. By integrating over this uncertainty, the Bayesian framework enhances model robustness and reduces sensitivity to hyperparameter values, offering a more principled alternative to MLE \citep{bayarri2007bayesian, Titsias2011mcmc}.

\begin{figure}[!ht]
    \centering
    \subfloat[Proposal Distribution: Uniform\label{fig:uni}]{\includegraphics[width=0.5\linewidth]{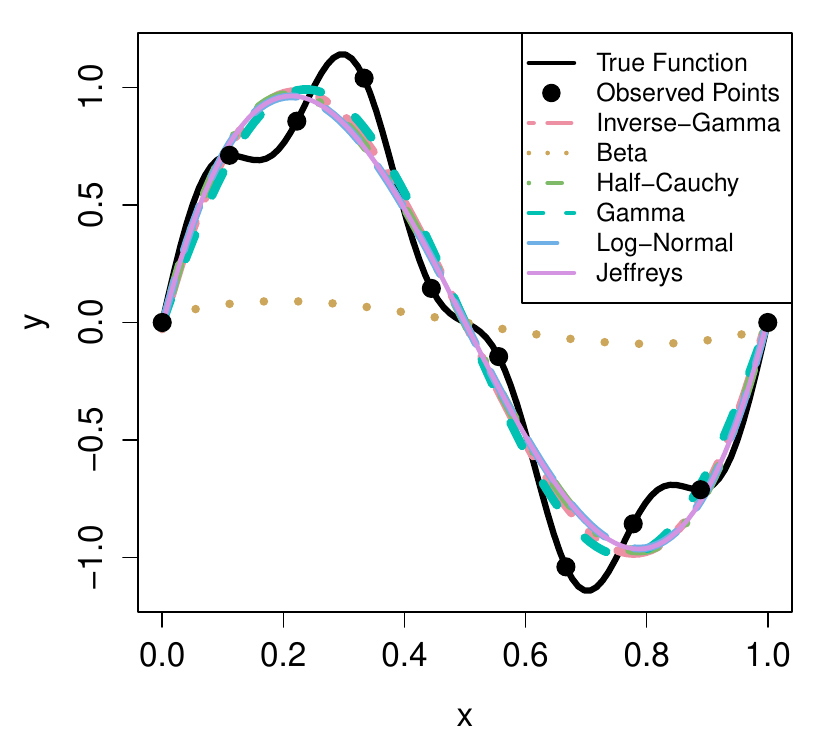}}
    \subfloat[Proposal Distribution: Normal\label{fig:normal}]{\includegraphics[width=0.5\linewidth]{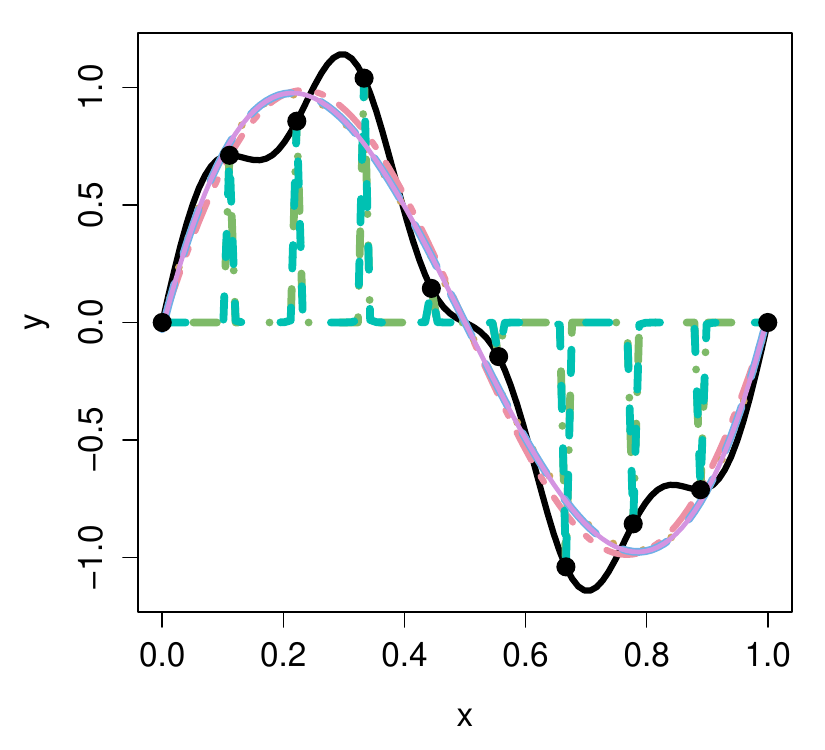}}
    \caption{Predictive performance of GPs on Higdon function under various priors for the lengthscale parameter, $\theta$. The black solid line represents the true underlying function, while black circular dots denote observed data points. Each colored curve corresponds to the predictive mean function obtained using a different prior for $\theta$, illustrating how prior and proposal choice influences model behavior.}
    \label{fig:higdon}
\end{figure}

\begin{figure}[!ht]
    \centering
    \subfloat[Proposal Distribution: Uniform\label{fig:uni-trace}]{\includegraphics[width=0.47\linewidth,height=0.45\linewidth]{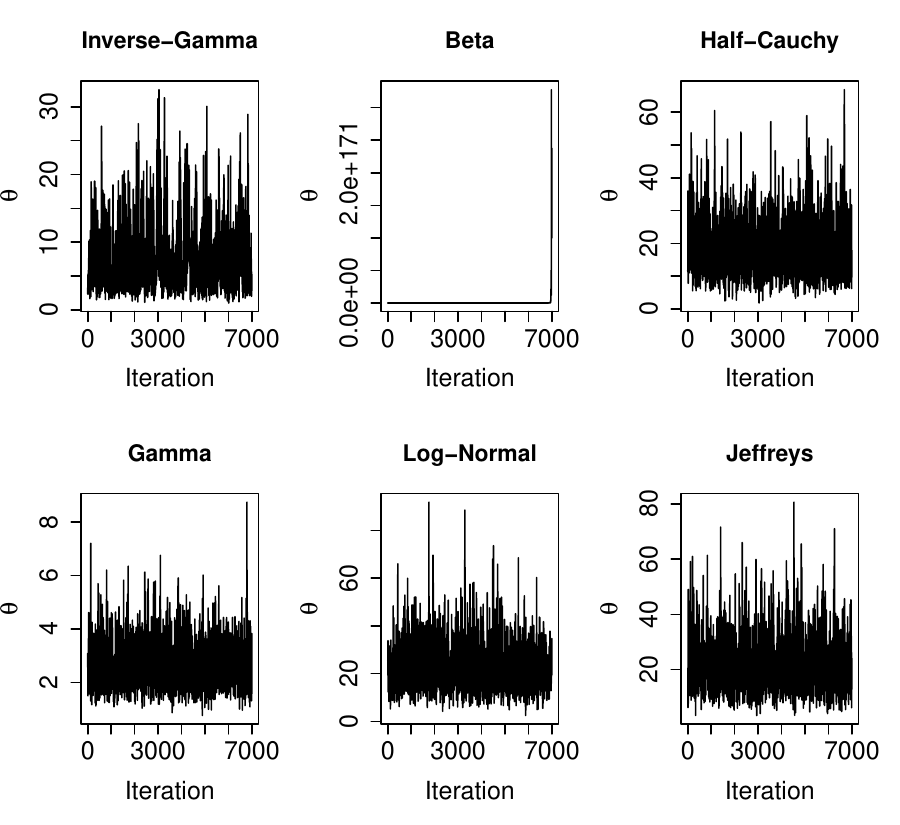}}
    \subfloat[Proposal Distribution: Normal\label{fig:normal-trace}]{\includegraphics[width=0.47\linewidth,,height=0.45\linewidth]{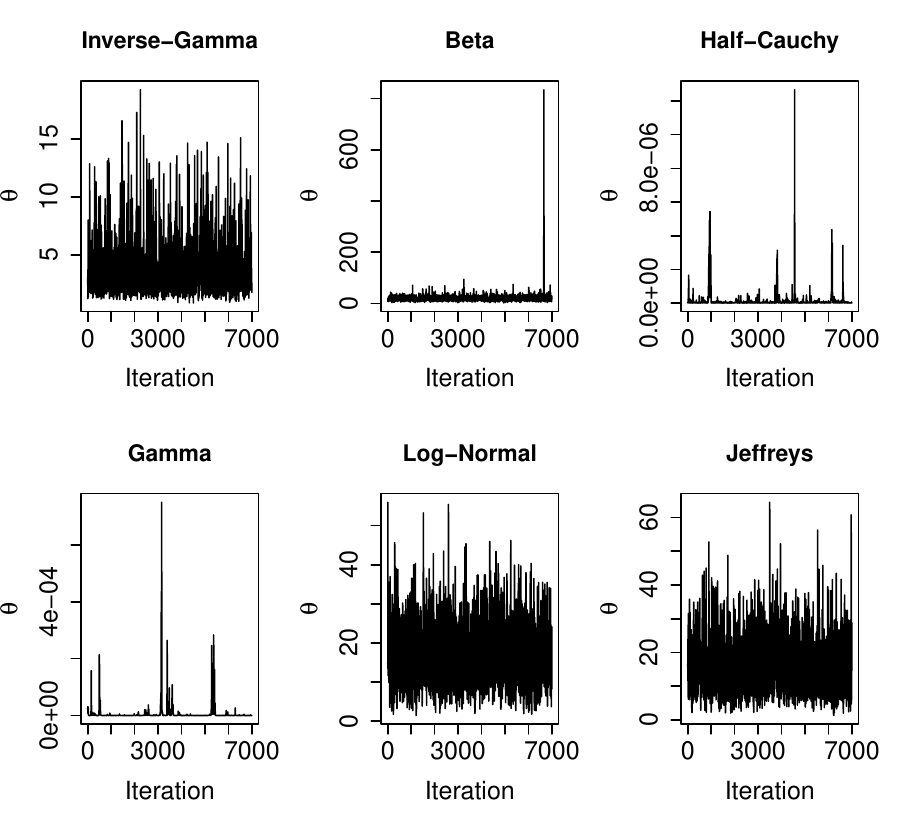}}
    \caption{Trace plots of the lengthscale parameter, $\theta$, under each prior distribution corresponding to the predictive models shown in \Cref{fig:higdon}. These plots illustrate how both the choice of prior and the proposal distribution influence the convergence characteristics. }
    \label{fig:higdon-trace}
\end{figure}

While the Bayesian approach often yields more appropriate and calibrated approximations, a key limitation is the need to specify prior distributions for the hyperparameters. In addition, the choice of proposal distribution and step size within the Metropolis-Hastings algorithm has a major impact on sampling efficiency and convergence \citep{roberts2001metropolis, robert1994mcmc, rosenthal2024proposal}. Although Metropolis-Hastings remains one of the most widely used MCMC methods, prior information is frequently unavailable or difficult to justify in practice, which can affect inference quality. To illustrate this, \Cref{fig:higdon} demonstrates how different prior and proposal choices for the lengthscale parameter, $\theta$, in GPs affect predictive performance on the one-dimensional Higdon function \citep{higdon2002space}, and \Cref{fig:higdon-trace} presents the corresponding trace plots for $\theta$ under each prior. The results suggest that the Beta prior may be unsuitable when paired with a uniform proposal distribution for this function. As shown in \Cref{fig:uni}, the predictive mean under the Beta prior fails to interpolate the data points, resulting in a flat function, and the trace plot in \Cref{fig:uni-trace} reinforces this observation. In contrast, predictive mean functions obtained under the other priors show a closer resemblance to one another. However, their trace plots indicate that the posterior distribution of $\theta$ varies in range depending on the prior. Under a normal proposal distribution, as shown in \Cref{fig:normal}, the Gamma and Half-Cauchy priors yield similar predictive means. They interpolate the training data points, but revert to the mean value in regions without data. Consistent with this, the trace plots exhibit a lack of convergence for these priors. Although the predictive performance of the Beta, Log-Normal, Inverse-Gamma, and Jeffrey priors appears comparable under the normal proposal distribution, \Cref{fig:normal-trace} shows that the posterior distribution from the Beta prior exhibits a sharp rise and fails to converge, distinguishing it from the others. Importantly, the step size in Metropolis-Hastings plays a critical role in controlling the stochasticity of the sampling process. Different step sizes can lead to significantly different behavior in both convergence and predictive performance. 

\subsection{Contributions}
Even though prior distributions play a crucial role in determining the reliability of GP inference, there has been relatively little theoretical or empirical guidance on how to select appropriate priors for practical applications. This lack of guidance is particularly problematic in scenarios with limited data, common in computer experiments, where the influence of the prior may be amplified. \cite{erickson2018comparison} provides a comparative overview of several GP software packages but does not investigate the role of priors. \cite{chen2018priors} examines the influence of different prior distributions used only to generate initial values for likelihood optimization within a frequentist framework. Since such priors serve solely as initialization schemes without considering genuine prior information, the predictive accuracy of GP regression remains nearly unaffected, although the optimized hyperparameters can vary. More recently, \cite{petit2023parameter} explored several selection criteria but mainly focused on the choice of the smoothness parameter $\nu$ in the Mat\'ern kernel from a frequentist perspective. In practice, several software packages acknowledge this issue and implicitly address it by assigning priors to hyperparameters, most commonly employing the Gamma, Half-Cauchy, or Log-Normal distributions \citep{GPflow, deepGP, GPBayes, NumPyro}. However, these implementations are typically limited to one or two default priors and rarely examine a broader range of prior choices or assess their impact on predictive performance.

To the best of our knowledge, this is the first work to systematically and empirically investigate how prior specifications influence GP hyperparameter inference and prediction within a Bayesian framework and to begin closing the aforementioned gap. Beyond priors themselves, we also examine the role of proposal distributions in Metropolis-Hastings sampling, which can have a substantial impact on the efficiency and convergence rate of the algorithm. Through extensive experiments across a range of priors and proposals, we demonstrate that some commonly used prior–proposal combinations lead to unstable posterior trajectories, poor mixing, or failure to interpolate even for smooth underlying functions. Our findings highlight practical scenarios in which certain priors or proposals may perform poorly and others may be more stable. Overall, these empirical results provide valuable insights for practitioners and underscore the importance of carefully considering both prior and proposal choices when applying Bayesian GP regression in practice.

% tentative 
The rest of this article is organized as follows. In Section \ref{sec:main}, we provide a review of GPs and fully Bayesian approaches to GP modeling. Numerical and real data studies are presented in Sections \ref{sec:studies} and \ref{sec:real}. Finally, Section \ref{sec:conclusion} concludes with a discussion.

\section{Bayesian GP Inference}\label{sec:main}
In this section, we begin by reviewing GPs and outline the structure of a hierarchical Bayesian GP model. We then describe the implementation of a Metropolis-within-Gibbs sampling algorithm for posterior inference. Finally, we introduce several commonly used priors for the covariance parameters and a selection of proposal distributions used in the Metropolis-Hastings updates.

\subsection{Gaussian processes}
Let $f:\mathbb{R}^d\rightarrow \mathbb{R}$ denote a blackbox function representing the expensive computer model. The input locations are collected in the matrix $\vect{X}_n = (\mathbf{x}_1, \ldots, \mathbf{x}_n)^\top \in \mathbb{R}^{n\times d}$, and $\vect{Y}_n\in \mathbb{R}^n$ denotes the corresponding output observations. A GP prior is placed on $f$, implying the multivariate normal distribution $\vect{Y}_n\sim \mathcal{N}(0, \tau^2\vect{K})$, where a zero mean is assumed for simplicity after centering the observations. Here, $\tau^2$ is a scale parameter, and $\vect{K}$ is the covariance matrix with each element $\vect{K}_{ij}=K(\mathbf{x}_i, \mathbf{x}_j)$ determined by a kernel function $K$. In this paper, the anisotropic squared exponential kernel is employed, defined as 
\begin{equation*}
    K(\mathbf{x},\mathbf{x}')=\exp{\left(-\sum^d_{i=1} \frac{\left(x_i-x_i'\right)^2}{\theta_i}\right)},% + g\mathbb{I}, % maybe because kernel function is without nugget g, but we add it after computing K. Please let me know if I'm wrong.
\end{equation*}
where $\boldsymbol \theta=(\theta_1, \ldots, \theta_d)$ are the lengthscale parameters, also referred to as covariance or kernel parameters, that control the strength of correlation along each input dimension. This kernel defines a GP that generates smooth functions, making it suitable for general-purpose learning tasks, such as those encountered in machine learning applications \citep{rasmussen2006gaussian}. To ensure numerical stability, we add a small jitter term $g$ to the diagonal of the kernel matrix \citep{ranjan2011computationally, peng2014choice}. Throughout this article, we assume $g$ is fixed to a non-zero positive value of $10^{-8}$.  Alternative kernel functions and their theoretical properties can be found in \cite{rasmussen2006gaussian} and \cite{stein1999interpolation}. 

Estimation of the covariance parameters $\boldsymbol \theta$ and $\tau^2$ is based on the log-likelihood of the observed outputs, given by 
\begin{equation}\label{eq:loglike}
    \log\mathcal{L}(\tau^2,\boldsymbol \theta) = -\frac{n}{2}\log 2\pi - \frac{n}{2}\log \tau^2 -    
    \frac{1}{2}\log|\vect{K}| - \frac{1}{2\tau^2}\vect{Y}_n^T\vect{K}^{-1}\vect{Y}_n . 
\end{equation}
For a given value of $\boldsymbol \theta$, the scale parameter $\tau^2$ that maximizes the log-likelihood in Equation~\eqref{eq:loglike} has the closed-form solution,
\begin{equation}\label{eq:tau2}
    \hat{\tau}^2(\boldsymbol \theta) = \frac{1}{n}\vect{Y}_n^T\vect{K}^{-}\vect{Y}_n .
\end{equation}
Substituting Equation~\eqref{eq:tau2} into Equation~\eqref{eq:loglike} yields the profile log-likelihood, which depends only on $\boldsymbol \theta$,
\begin{equation}\label{eq:profile}
    \log\mathcal{L}(\boldsymbol \theta) = c - \frac{n}{2}\log \vect{Y}_n^T\vect{K}^{-}\vect{Y}_n - \frac{1}{2}\log |\vect{K}|,
\end{equation}
where $c$ is a constant independent of $\boldsymbol \theta$. As a result, statistical inference for the GP relies entirely on the estimation of $\boldsymbol \theta$. The remainder of this article investigates how different prior choices for the lengthscale parameters $\boldsymbol \theta$ affect predictive performance. 

To extend this framework, we consider a hierarchical Bayesian GP model where the latent function $f$ is explicitly modeled and the covariance parameters are treated probabilistically. Given observed data $(\vect{X}_n, \vect{Y}_n)$, where each $Y_l$ is a noisy realizations of the latent function $f$, modeled as $Y_l=f_l+\epsilon_l$ where $\epsilon_l\sim (0, \eta^2)$, with Gaussian noise $\epsilon_l$ having variance $\eta^2$. The hierarchical GP model is defined as 
\begin{align*}
    &\text{Prior over covariance parameter: } \boldsymbol \theta\sim p(\boldsymbol \theta) \notag \\
    &\text{Prior over latent function: } \vect{f}\mid \vect{X}_n, \boldsymbol \theta \sim N(0, \tau^2 \vect{K}) \\
    &\text{Likelihood of observed data: } \vect{Y_n}\mid \vect{f} \sim N(\vect{f}, \eta^2\mathbb{I}). \notag
\end{align*}
where $\vect{f}=(f_1,\dots, f_n)$. 
The joint posterior over unknown parameters given by 
\begin{equation*}
    p(\vect{f}, \boldsymbol \theta\mid \vect{Y_n}) = \frac{p(\vect{Y_n}\mid \vect{f})\,p(\vect{f}\mid \boldsymbol \theta)\,p(\boldsymbol \theta)}{Z}
\end{equation*}
where $Z$ is the normalization constant. 

Our objective is to predict the output $Y(\vect{x}_{\text{new}})$ at an unobserved location $\vect{x}_{\text{new}}\in\mathcal{X}$, where $\mathcal{X}$ denotes an $m\times d$ matrix of new input locations. The GP predictive distribution of $Y(\mathcal{X})$ given $(\vect{X}_n, \vect{Y}_n)$ is multivariate normal, $Y(\mathcal{X})|(\vect{X}_n, \vect{Y}_n) \sim \mathcal{N}(\vect{\mu}^*, \Sigma^*)$, with
\begin{align}
    \vect{\mu}^*(\mathcal{X}) & = \vect{K}(\mathcal{X}, \vect{X}_n)\vect{K}^{-1}\vect{Y}_n \label{eq:pred-mean}\\
    \Sigma^*(\mathcal{X}) &= \tau^2(\vect{K}(\mathcal{X},\mathcal{X})-\vect{K}(\mathcal{X}, \vect{X}_n)\vect{K}^{-1}\vect{K}(\vect{X}_n,\mathcal{X})) \ ,\ \label{eq:pred-cov}
\end{align}
where $\vect{K}(\mathcal{X}, \vect{X}_n)$ is the $m\times n$ covariance matrix between the new and training locations.

\subsection{Bayesian implementation} \label{sec:implementation}
To perform full Bayesian inference, we adopt a hierarchical framework in which the covariance parameters, $\boldsymbol{\theta}$ are treated as random variables with specified prior distributions. The posterior inference is carried out using a Metropolis-within-Gibbs sampling algorithm, where the Metropolis-Hastings step is used to update each component of $\boldsymbol \theta$. Specifically, our objective is to examine how different prior choices for $\boldsymbol \theta$ affect predictive performance. In our analysis, a zero-mean GP is assumed, a closed-form update is used for $\tau^2$, and the jitter $g$ is fixed to a small value to ensure numerical stability. Before detailing the implementation, the principles of Gibbs sampling and Metropolis-Hastings are briefly reviewed. 

Gibbs sampling, originally proposed by \cite{gelfand1990gibbs}, is an MCMC method that iteratively generates samples from the full conditional distributions of each parameter. Given a target distribution $p(\boldsymbol \theta)=p(\theta_1,\ldots, \theta_d)$, the Gibbs sampler successively updates each component $\theta_i$ by drawing from its conditional distribution $p(\theta_i \mid \theta_1, \ldots, \theta_{i-1}, \theta_{i+1}, \ldots, \theta_d)$. The resulting Markov chain converges to $p(\boldsymbol \theta)$ under mild regularity conditions \citep{robert1994mcmc}. Hence, after a sufficiently large number of iterations $N$, the sequence $(\boldsymbol \theta^{(0)}, \ldots, \boldsymbol \theta^{(N)})$ can be regarded as approximate realizations from the target distribution. A key limitation of Gibbs sampling is that convergence may be slow, particularly when the full conditional distributions are not available in closed form or are expensive to sample from directly. 

The Metropolis-Hastings algorithm, introduced by  \citet{metropolis1953} and \citet{hastings1970}, offers a more flexible alternative. Instead of relying on the full conditional distributions, it utilizes a proposal distribution $q(\theta_i^* \mid \theta_i^{(t-1)})$ to suggest a candidate value $\theta_i^*$ for the $i$th component of the current state $\boldsymbol \theta^{(t-1)}$. The flexibility of Metropolis-Hastings lies in the choice of the proposal distribution and tuning of the acceptance rate, both of which can significantly affect the efficiency of the sampler. 

% create algorithm??????
In the Metropolis-within-Gibbs update for $\theta_i$, an initial value $\boldsymbol \theta^{(0)}$ is selected and the corresponding scale parameter $\tau^2$ is computed using Equation~\eqref{eq:tau2}. At iteration $t$, the Markov chain is at the current state $\boldsymbol \theta^{(t-1)}$, and the likelihood in Equation~\eqref{eq:profile} is evaluated at $\boldsymbol \theta^{(t-1)}$. A candidate proposal $\theta_i^*$ is then generated from $q(\theta_i^* \mid \theta_i^{(t-1)})$, and the prior density $p(\boldsymbol \theta^{*})$ and the likelihood $\mathcal{L}(\boldsymbol \theta^*)$ are evaluated. The acceptance probability is then calculated as:
\begin{equation*}
    \alpha=\min\left(1, \frac{p(\boldsymbol \theta^*)\mathcal{L}(\boldsymbol \theta^*)}{p(\boldsymbol \theta^{(t-1)})\mathcal{L}(\boldsymbol \theta^{(t-1)})}\times \frac{q(\theta_i^{(t-1)} \mid \theta_i^*)}{q(\theta_i^* \mid \theta_i^{(t-1)})}\right),
\end{equation*}
where $\boldsymbol \theta^*$ denotes the vector obtained by replacing $\theta_i^{(t-1)}$ with $\theta_i^*$ in $\boldsymbol \theta^{(t-1)}$, while keeping all other components $\{\theta_j:j\neq i \}$ fixed. If the proposal is accepted, we set $\boldsymbol \theta^{(t)}=\boldsymbol \theta^*$; otherwise we retain the previous value, $\boldsymbol \theta^{(t)}=\boldsymbol \theta^{(t-1)}$. This Metropolis step is applied sequentially for $i=1,\ldots,d$, yielding one full Gibbs sweep for iteration $t$. After $N$ iterations, we obtain posterior samples $\{\boldsymbol \theta^{(t)}\}^N_{t=1}$ and corresponding scale parameter $\{\tau^{2(t)}\}^N_{t=1}$, where each $\tau^{2(t)}$ is computed based on the sampled  $\boldsymbol \theta^{(t)}$ value. 

% initial value details
The selection of the initial value $\boldsymbol \theta^{(0)}$ is important for ensuring efficient convergence of the MCMC sampler. \cite{basak2021initialvalue} introduces a practical approach for initializing parameters based on the empirical moments of the data. Specifically, they suggest initializing each lengthscale parameter with the sample standard deviation of the corresponding input dimension $\mathbf{x}_{(i)}=(x_{1i},\dots, x_{ni})^\top$, which reflects the distance over which meaningful variation occurs in the input space \citep{rasmussen2006gaussian}. Following this strategy, we initialize each component of $\boldsymbol \theta^{(0)}$ with the sample standard deviation of $\mathbf{x}_{(i)}$ in our MCMC procedure. As discussed in Section \ref{sec:intro}, both the prior and proposal distributions play a crucial role in determining the predictive performance of the model. The specific choices considered in our analysis are described in the following section. 

Given MCMC samples $\{\boldsymbol \theta^{(t)}\}_{t=B+1}^N$ obtained after discarding the first $B$ burn-in iterations, we compute the corresponding predictive mean $\mu^{* (t)}$ (Equation~\eqref{eq:pred-mean}) and covariance $\Sigma^{* (t)}$ (Equation~\eqref{eq:pred-cov}) for each retained iteration $t$. The posterior predictive mean and covariance are then obtained as 
\begin{align*}
    \mu^* &= \frac{1}{N-B}\sum^{N}_{t=B+1}\mu^{*(t)}, \\
    \Sigma^* &= \frac{1}{N-B} \sum^{N}_{t=B+1}\Sigma^{*(t)} + \mathrm{Cov}\left(\left\{ \mu^{*(t)} \right\}_{t=B+1}^N\right),
\end{align*}
where the first term in $\Sigma^*$ represents aleatoric uncertainty, and the second term reflects epistemic uncertainty arising from posterior variability in the model parameters. This decomposition follows from the law of total variance.

\subsubsection{Prior distributions}
Although the influence of the prior tends to diminish with large datasets, it can become substantial in limited data scenarios, common in computer experiments, where each observation is costly and sample sizes are small \citep{basak2021initialvalue}, and in environmental modeling, where data collection is often constrained by ecological factors \citep{wesner2021ecological}. In such cases, careful specification of the prior distribution is essential for reliable inference. This is particularly true for the lengthscale parameter in GPs, as it controls the smoothness of the response surface and directly affects model flexibility. Several prior choices have been used for the lengthscale parameter $\theta_i$, each with different implications for regularization and predictive behavior. In this work, we investigate six such priors, focusing on those implemented in widely used software packages. 

Stan \citep{stan} is a probabilistic programming language for specifying and fitting Bayesian statistical models. Its Stan User's Guide illustrates the use of the Inverse-Gamma distribution, $\text{IG}(\alpha, \beta)$, as an example prior for the lengthscale parameter $\theta_i$, with parameters $\alpha=5$ and $\beta=5$. However, as noted by \cite{berger2001reference}, such vague yet proper priors can be highly sensitive to the specific choices of $\alpha$ and $\beta$. 

The \texttt{GPBayes} package in R \citep{GPBayes} provides users with several options for specifying prior distributions for $\theta_i$, including the Beta, Half-Cauchy, and reference priors. Under the default settings, the Beta distribution, $\text{Beta}(\alpha, \beta)$, is used with parameters $\alpha=1$ and $\beta=1$, corresponding to a uniform prior over the unit interval. While the package implements the reference prior, it is limited to the isotropic settings explored by \cite{berger2001reference}. An extension to anisotropic scenarios is provided by \cite{paulo2005bayes}, who show that the resulting posterior remains proper under certain regularity conditions. We include both Beta and Half-Cauchy priors, $\text{Half-Cauchy}(\sigma)$, from this package in our analysis.

The \texttt{deepgp} package \citep{deepGP} in R employs a Gamma prior, $\text{Gamma}(\alpha, \beta)$, for $\theta_i$. Its default settings specify a shape parameter of $\alpha=1.5$ and a rate of $\beta=3.9/1.5$. Further details are provided in \cite{sauer2023active}. Similarly, the \texttt{GPflow} library in Python also adopts a Gamma prior for $\theta_i$ \citep{GPflow}. Since the Gamma prior can be sensitive to the choice of its shape and rate parameters, careful tuning is often necessary to avoid overly informative or diffuse priors. 

The \texttt{NumPyro} library in Python \citep{NumPyro} allows users to define custom priors for model parameters, including distributions such as Log-Normal, Gamma, and Half-Cauchy. An example from its documentation demonstrates the use of a Log-Normal prior, $\text{Lognormal}(\mu, \sigma^2)$, with $\mu=0$ and $\sigma=10$. This choice reflects a broad, weakly informative prior that allows for a wide range of plausible values. We explore the Log-Normal prior in our comparison.

The Jeffreys prior and reference prior are both designed to be non-informative and are discussed in detail by \cite{berger2001reference} and \cite{paulo2005bayes}. The Jeffreys prior, originally proposed by \cite{jeffrey1946}, is defined over the parameter space with density proportional to the square root of the determinant of the Fisher Information matrix, $p(\boldsymbol \theta)\propto |I(\boldsymbol \theta)|^{1/2}$. For the lengthscale parameter in GPs, this prior takes the form: 
\begin{align*}
    p(\boldsymbol \theta) \propto \left| \mathbf{S}-\frac{1}{n} \mathbf{t} \mathbf{t}^\top \right|^{\frac{1}{2}}
\end{align*}
where the $t_i=\mathrm{tr}\left(\mathbf{K}^{-1} \frac{\partial \mathbf{K}}{\partial \theta_i}\right)$ and $\mathbf{S}=\left(S_{ij}\right)_{i,j=1}^d=\mathrm{tr}\left(\mathbf{K}^{-1} \frac{\partial \mathbf{K}}{\partial \theta_i} \mathbf{K}^{-1} \frac{\partial \mathbf{K}}{\partial \theta_j}\right)_{i,j=1}^d$. The full derivation is provided in the Appendix~\ref{appendix}. This prior is often used when minimal prior information is available. As discussed above, the reference prior is implemented in the \texttt{GPBayes} package. In this study we investigate the effect of the Jeffreys prior on model performance, noting that \cite{paulo2005bayes} demonstrate its propriety under specific conditions. 

To ensure a comprehensive evaluation of prior choices for the lengthscale parameter in GPs, we focus on six distributions, Inverse-Gamma, Beta, Half-Cauchy, Gamma, Log-Normal, and Jeffreys priors. These priors are implemented in various software packages and represent the most commonly used approaches in practice. The hyperparameters for each distribution are set as follows: $\alpha=5$ and $\beta=5$ for the Inverse-Gamma, $\alpha=1$ and $\beta=1$ for the Beta, scale parameter $\sigma=1$ for the Half-Cauchy, $\alpha=1.5$ and $\beta=3.9/1.5$ for the Gamma, and mean $\mu=0$ and standard deviation $\sigma=10$ for the Log-Normal. These settings follow the defaults provided in respective software packages or user guide.

\subsubsection{Proposal distributions}\label{sec:proposal}
While the selection of the prior distribution is important, the choice of proposal distribution in Metropolis-Hastings sampling also influences sampling efficiency and posterior exploration. The choice is highly application-dependent \citep{brooks2011mcmc, roberts2001metropolis}. One proposal that works well for one target distribution may be extremely ineffective for another. In our analysis, we examine two widely used proposal distributions, uniform and normal, and assess their behavior in the context of GPs. 

To ensure positivity of the lengthscale parameter $\theta_i$, we adopt a multiplicative random walk proposal \citep{andrieu2008proposal}. Specifically, proposals are drawn from the distribution $\theta_i^* \sim \text{Unif}\left(\theta_i^{(t-1)}/u, u\theta_i^{(t-1)}\right)$, where $u>1$ serves as a multiplicative step-size tuning parameter. This formulation maintains both log-symmetry and positivity, %It is computationally similar to the Random Dive Metropolis-Hastings algorithm introduced by \cite{dutta2012multiplicative}, and 
and is implemented in the \texttt{deepgp} package. The tuning parameter $u$ significantly affects both the efficiency and numerical stability of the sampler. For instance, \cite{steven2025monotonic} employ $u=2$, which allows the sampler to explore a wider range of values around the current state. However, we observed that using $u=2$ in conjunction with a Beta prior can lead to numerical instability in higher-dimensional settings, as the proposal value may diverge to infinity. To mitigate this behavior, we set $u=2$ for lower-dimensional examples and adopt smaller values of $u$ such as $u=1.5$ or $u=1.2$ in higher-dimensional cases throughout our numerical experiments.

% normal
An alternative proposal strategy involves a Gaussian random walk defined as $\theta_i^* \sim N(\theta_i^{(t-1)}, \sigma^2)$ \citep{Gelman1997proposal, rosenthal2024proposal}. To ensure that the proposed values of $\theta_i$ remain strictly positive, we perform updates in the log domain. Specifically, we sample $\log(\theta_i^*) \sim N(\log(\theta_i^{(t-1)}), \sigma^2)$, and then set $\theta_i^*=\exp(\log(\theta_i^*))$. This transformation guarantees positivity, and the acceptance ratio includes a Jacobian correction $\big|\frac{\partial \theta_i^*}{\partial \log{(\theta_i^*)}} \big|=\theta_i^*$ to account for the change of variables. In addition, to prevent numerical issues near zero, we utilize a truncated normal proposal with a lower bound of $10^{-10}$. In contrast to the multiplicative proposal, here the tuning parameter is the standard deviation $\sigma$, which controls the scale of additive perturbations. Larger values, such as $\sigma=0.5$, may lead to instability or low acceptance rates, particularly when combined with a Beta prior in high-dimensional settings. Consistent with the uniform proposal strategy, we adjust $\sigma$ based on dimensionality, using $\sigma=0.5$ for lower-dimensional cases and smaller values for higher-dimensional settings. 

The interaction between prior and proposal distributions is critical to the efficiency and stability of MCMC sampling in GP models. The following sections present both simulation and real-data results that illustrate these effects.

\section{Numerical Studies}\label{sec:studies}

In this section, we present a series of numerical experiments designed to evaluate how different prior specifications influence the performance of GP models. In particular, we investigate six priors--Inverse-Gamma, Beta, Half-Cauchy, Gamma, Log-Normal, and Jeffreys--each combined with the proposal distribution discussed in Section~\ref{sec:proposal}. The evaluation is conducted using four test functions defined as follows:
\begin{align*}
\begin{cases}
 \quad f(x) &= \sin{\frac{2\pi x}{10}}+0.2\sin{\frac{2\pi x}{2.5}}, \quad x\in [0,10], \\
 \quad f(\mathbf{x}) &= -\sum^4_{i=1}\alpha_i \exp{\left(-\sum^3_{j=1}A_{ij}(x_j-P_{ij})^2\right)}, \quad \mathbf{x}=(x_1,x_2,x_3)\in [0,1]^3, \\
 \quad f(\mathbf{x}) &= 100(x_1^2-x_2)^2+(x_1-1)^2+(x_3-1)^2+90(x_3^2-x_4)^2+10.1((x_2-1)^2+(x_4-1)^2)\\
&+19.8(x_2-1)(x_4-1), \quad \mathbf{x}=(x_1,x_2,x_3,x_4)\in [-10,10]^4, \\
 \quad f(\mathbf{x}) &=
\frac{2\pi T_u (H_u - H_l)}
{\log(r/r_w)\left(1 + \frac{2LT_u}{\log(r/r_w)r_w^2K_w} + \frac{T_u}{T_l}\right)},
\end{cases}
\end{align*}
where $\alpha=(1.0, 1.2, 3.0, 3.2)^T$, $A=\begin{pmatrix}
    3.0&10&30\\
    0.1&10&35\\
    3.0&10&30\\
    0.1&10&35
\end{pmatrix}$, $P=10^{-4}\begin{pmatrix}
    3689&1170&2673\\
    4699&4387&7470\\
    1091&8732&5547\\
    381&5743&8828
\end{pmatrix}$,
and the input variables of the fourth function lie in $r_w\in[0.05, 0.15]$, $r\in[100, 50000]$, $T_u\in[63070, 115600]$, $H_u\in[990, 1110]$, $T_l\in[63.1, 116]$, $H_l\in[700, 820]$, $L\in[1120, 1680]$, and $K_w\in[9855, 12045]$. The first synthetic function is the Higdon function from \cite{higdon2002space}. The second example is the three-dimensional function proposed by \cite{dixon1978hartmann}, commonly referred to as the Hartmann 3-d function. The third is the Colville function \citep{song2019radial, rahnamayan2007novel}. The last synthetic example is the Borehole function, originally introduced by \cite{morris1993bayesian} and \cite{worley1987deterministic}. 

For each test function, the training and testing points are generated using Latin hypercube sampling \citep{stein1987large}. The training design consists of $10d$ points, following the guideline of \cite{loeppky2009choosing}, while the testing design includes $100d$ points to ensure reliable assessment of predictive performance, where $d$ denotes the input dimension of the function. Prior to analysis, the response variable is centered to have a zero mean.

We obtain the posterior distribution for $\boldsymbol \theta$ through an MCMC process, alongside the scale parameter $\tau^2$, which is estimated by solving Equation~\ref{eq:tau2}. The total number of MCMC iterations depends on the dimensionality of test function, with $N=10000d$. To ensure convergence and reduce the influence of initial values, the first $30\%$ of samples are discarded as burn-in. The remaining $70\%$ of posterior samples are then used to compute predictive quantities, as described in Section~\ref{sec:implementation}, by averaging the predictive mean and aggregating posterior uncertainty. 

The entire process is repeated 100 times, and predictive performance is assessed using three criteria: root mean squared error (RMSE) for point-prediction accuracy, the continuous ranked probability score (CRPS) for the quality of the full predictive distribution \citep{gneiting2007strictly, bastos2009diagnostic}, and the 95\% predictive interval coverage rate (PICR) for uncertainty quantification. The metrics are defined as follows:
\begin{align*}
\text{RMSE} &= 
\sqrt{\frac{1}{n_\text{test}}\sum_{i=1}^{n_\text{test}} (y_i - \hat{\mu}_i)^2}, \\
\text{CRPS} &= \frac{1}{n_\text{test}}\sum_{i=1}^{n_\text{test}} \hat{\sigma}_i \left(2\phi(z_i) + z_i \left(2\Phi(z_i) - 1\right) -\frac{1}{\sqrt{\pi}} \right), \quad \text{for} \quad z_i = \frac{y_i - \hat{\mu}_i}{\hat{\sigma}_i}, \\
\text{PICR} &= 
\frac{1}{n_\text{test}}\sum_{i=1}^{n_\text{test}} 
\mathbb{I}
\left(
\hat{\mu}_i - z_{0.975} \hat{\sigma}_i \le y_i \le 
\hat{\mu}_i + z_{0.975} \hat{\sigma}_i
\right),
\end{align*}
where $\hat{\mu}_i$ and $\hat{\sigma}_i^2$ denote the predictive mean and variance at the $i$th test point, respectively, $\phi(\cdot)$ and $\Phi(\cdot)$ are the standard normal probability density and cumulative distribution functions, and $z_{0.975}$ denotes the 97.5th percentile of the standard normal distribution. Smaller RMSE and CRPS values indicate better predictive performance. For PICR, achieving a coverage rate close to the nominal 95\% is ideal.  

\begin{figure}[!ht]
    \centering
    \text{RMSE}\\
    \subfloat[Uniform\label{fig:3a}]{\includegraphics[width=0.43\linewidth, height=0.33\linewidth]{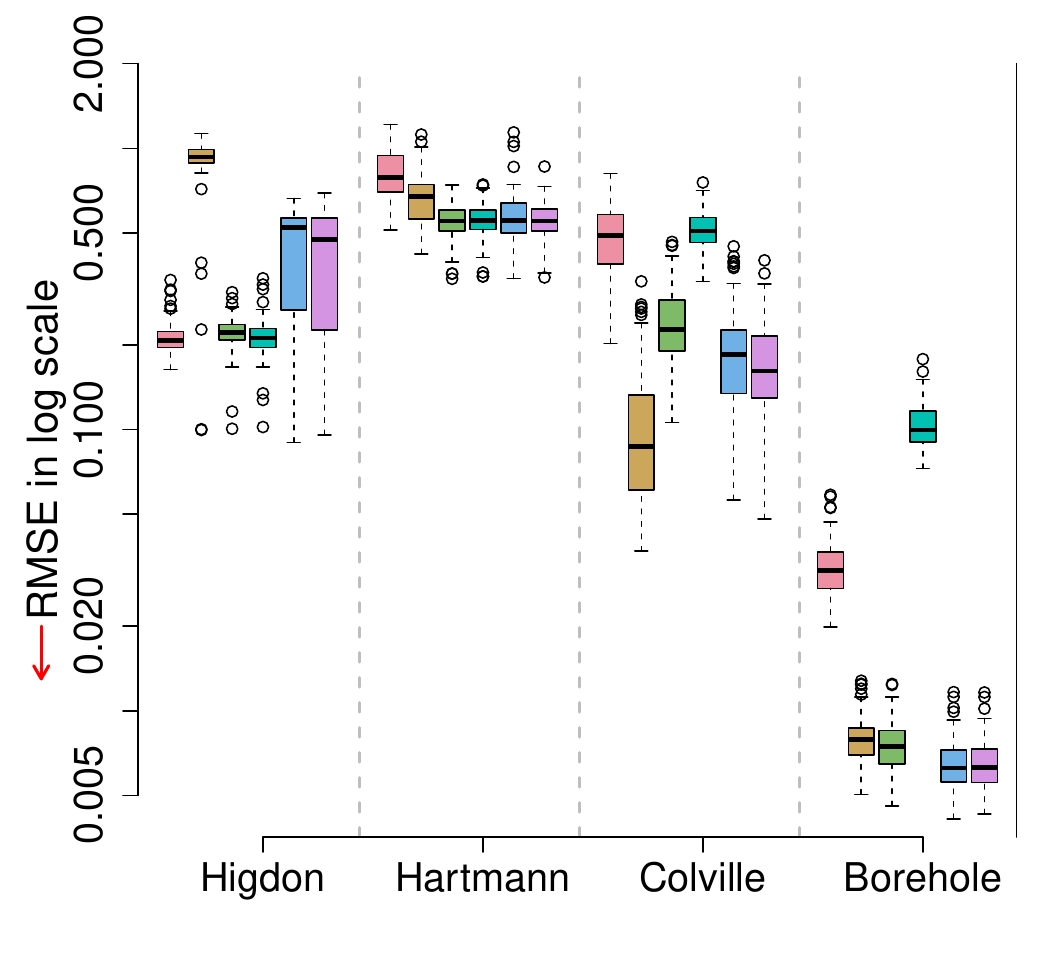}} 
    \subfloat[Normal\label{fig:3b}]{\includegraphics[width=0.43\linewidth, height=0.33\linewidth]{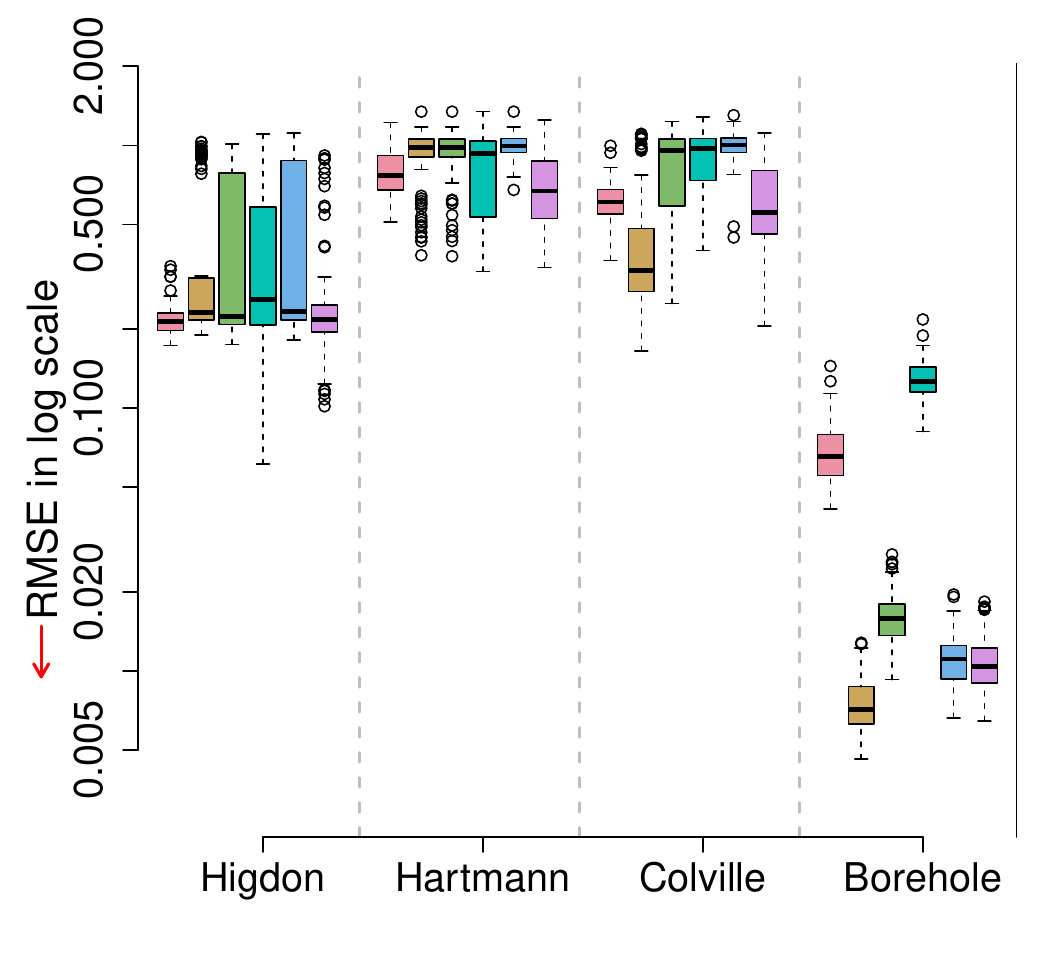}} \\
\text{CRPS}\\
    \subfloat[Uniform\label{fig:3c}]{\includegraphics[width=0.43\linewidth, height=0.33\linewidth]{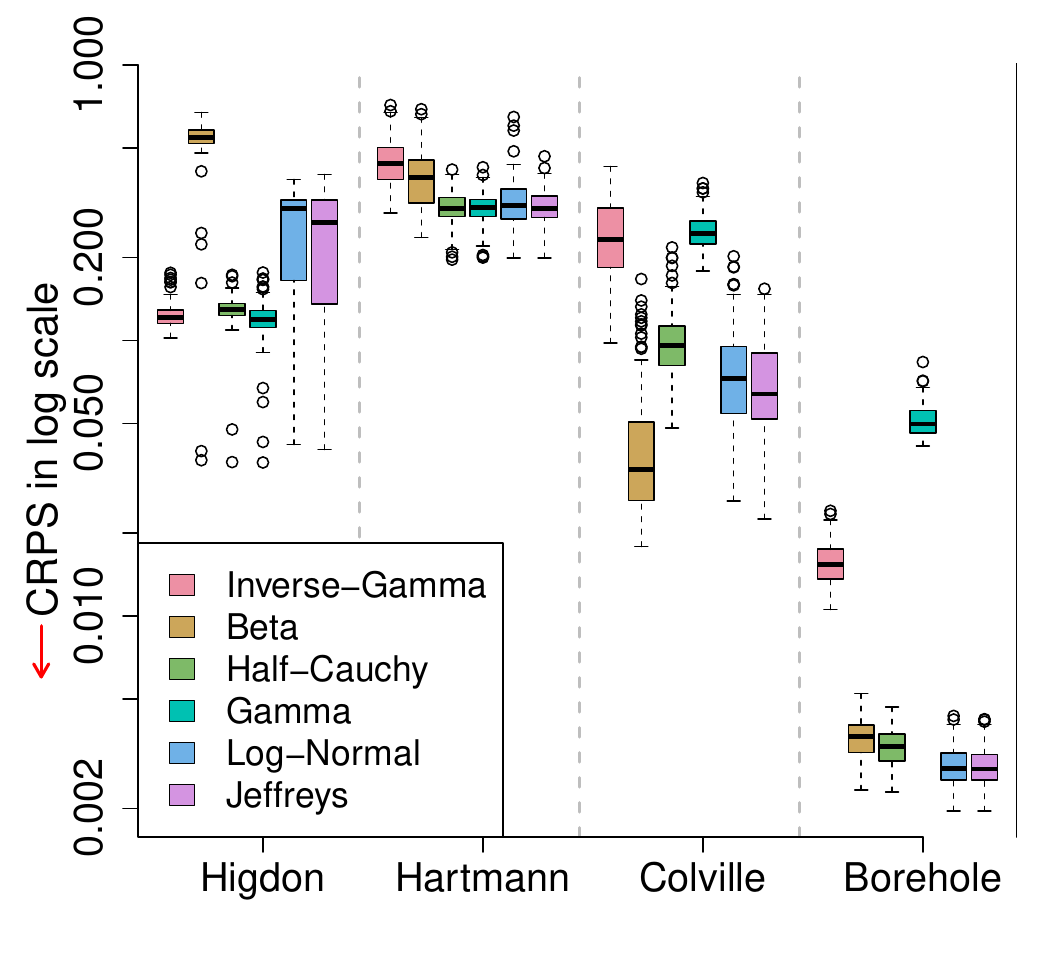}} 
    \subfloat[Normal\label{fig:3d}]{\includegraphics[width=0.43\linewidth, height=0.33\linewidth]{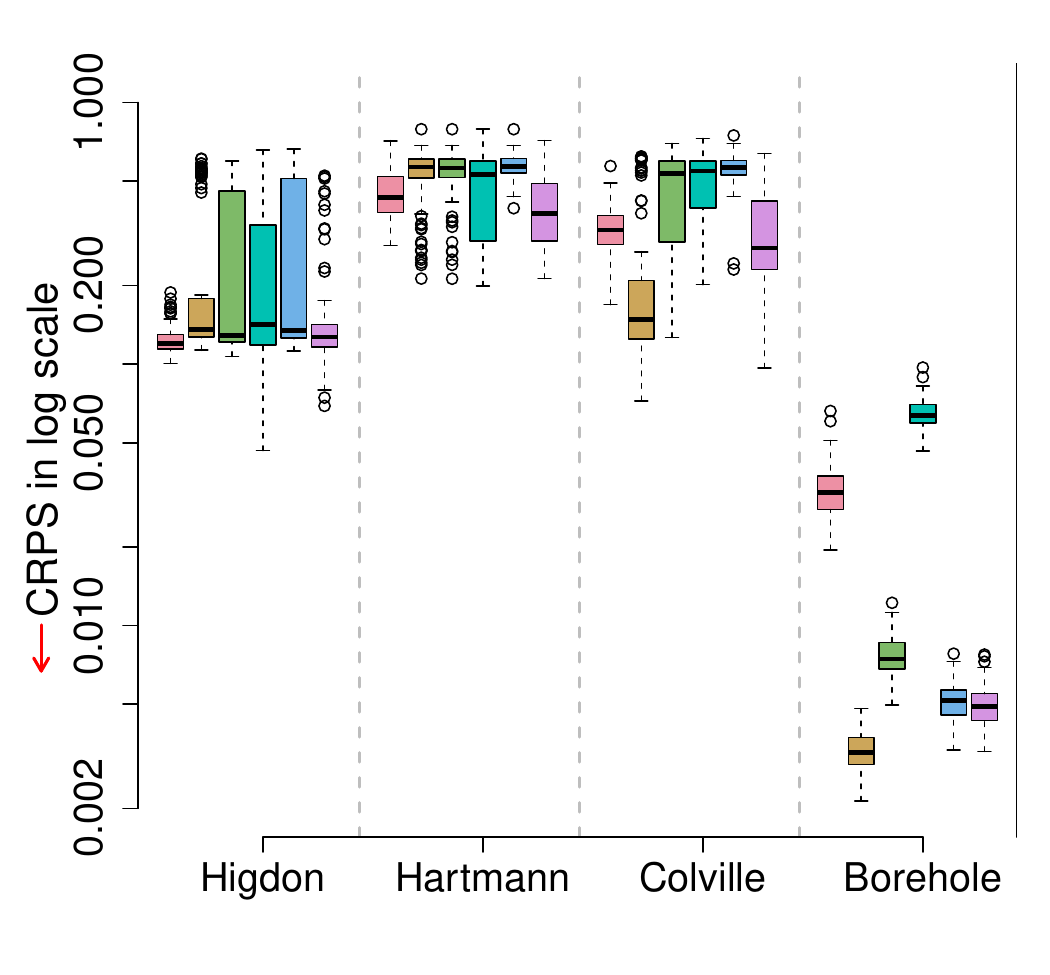}} \\
\text{PICR}\\
    \subfloat[Uniform\label{fig:3e}]{\includegraphics[width=0.43\linewidth, height=0.33\linewidth]{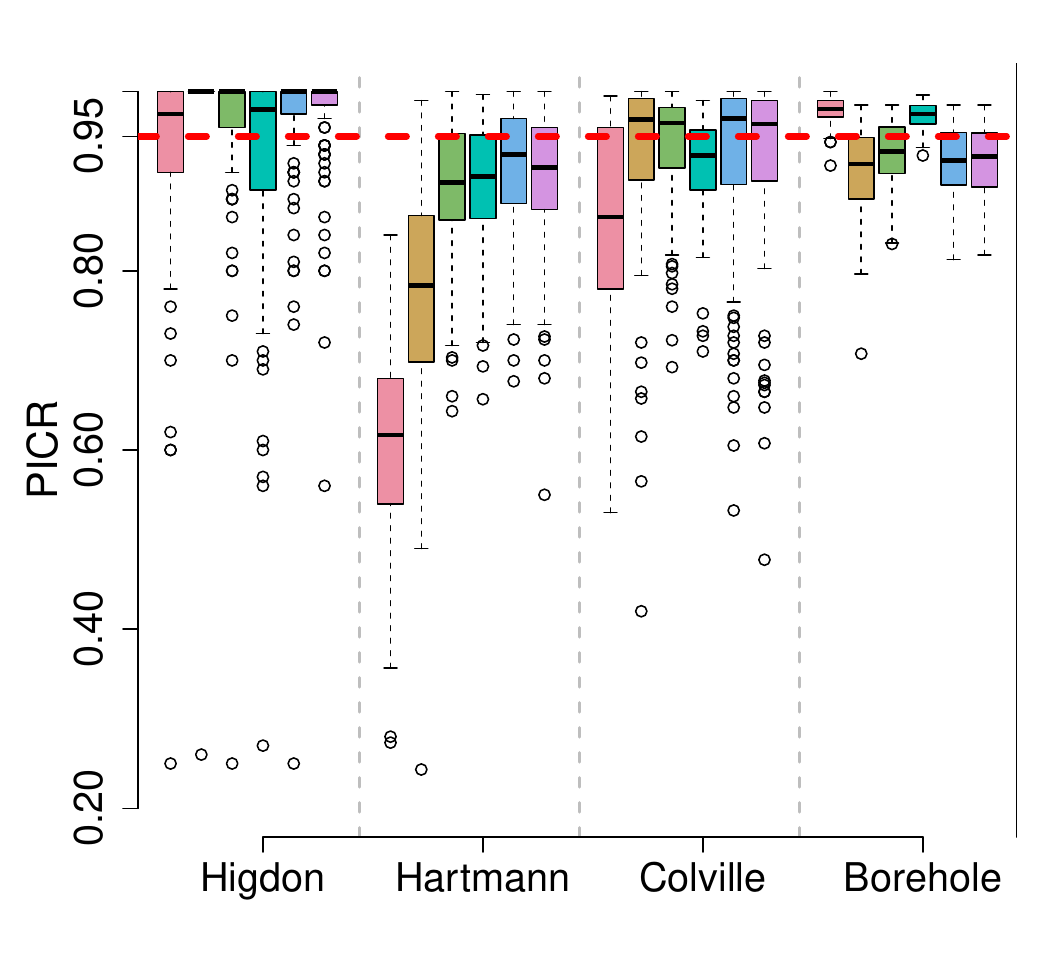}} 
    \subfloat[Normal\label{fig:3f}]{\includegraphics[width=0.43\linewidth, height=0.33\linewidth]{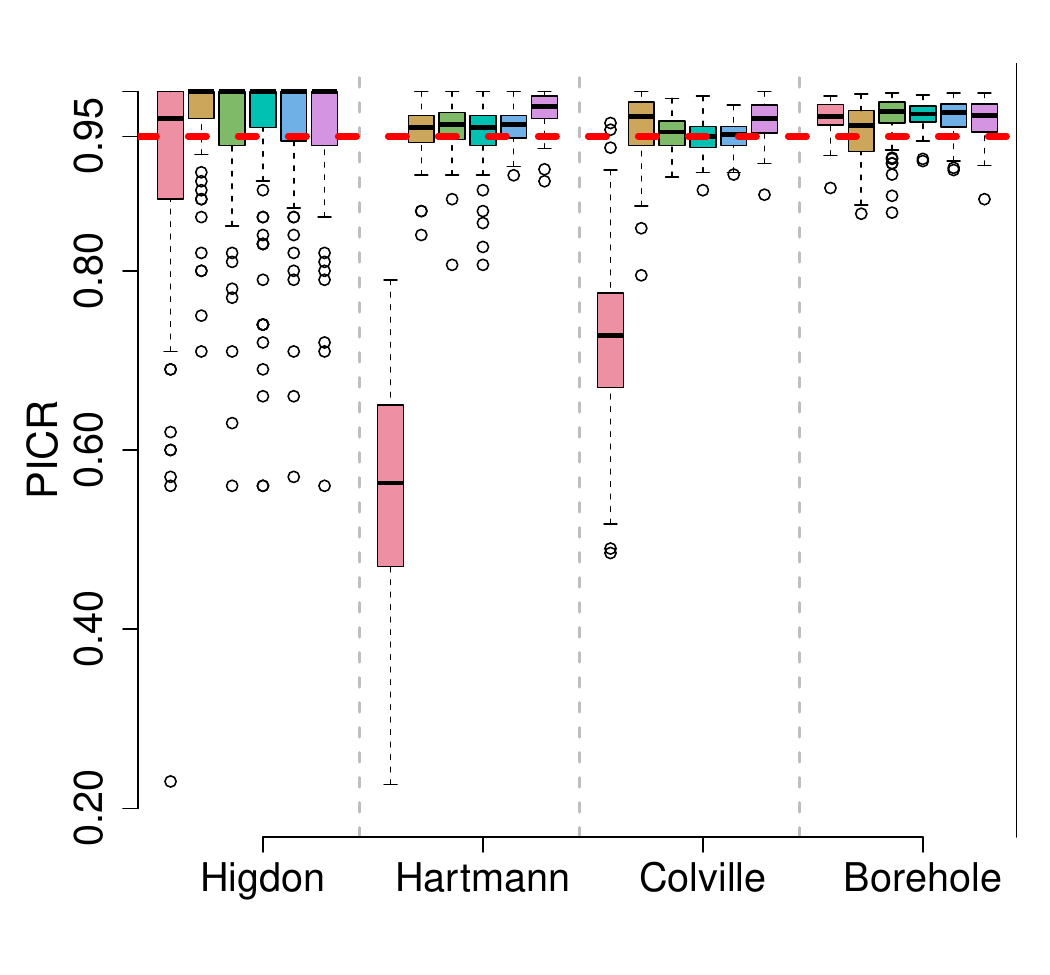}}
    \caption{RMSE (upper), CRPS (middle), and PICR (lower) across 100 repetitions for the synthetic functions using a uniform proposal distribution (left) and a normal proposal distribution (right).}
    \label{fig:simulation}
\end{figure}

\Cref{fig:simulation} summarizes the results of the analysis. As illustrated in the figures, both the choice of prior and proposal distributions significantly influence predictive performance in GPs. Notably, the predictive performance under the Beta prior shows significant variation depending on the choice of proposal distribution, reflected in RMSE and CRPS for the Higdon function, and in PICR for the Hartmann function. Compared to other priors, the Beta prior exhibits greater variability in performance depending on the proposal strategy. For the Borehole function, both the Inverse-Gamma and Gamma priors tend to underperform in terms of RMSE and CRPS, while yielding relatively higher PICR values than other priors. This behavior can be attributed to their tendency to produce smaller mean posterior values for $\boldsymbol \theta$, given the hyperparameter settings of $\alpha=5$, $\beta=5$ for the Inverse-Gamma, and $\alpha=1.5$ and $\beta=3.9/1.5$ for the Gamma prior. Smaller values of $\boldsymbol \theta$ correspond to shorter correlation lengths, which in turn lead to rougher predictive surfaces and potentially overfit models. We also observed that modifying these hyperparameter values results in markedly different predictive behaviors, highlighting the sensitivity of these priors to their parameterization. This underscores the importance of careful selection of prior hyperparameters when modeling with GPs.

%Maybe write something about how different priors affect to the smoothness of GP? How can we assess MCMC values? in terms of true value

% https://ww2.mathworks.cn/help/pde/ug/electrostatic-potential-in-air-filled-frame-femodel.html
\section{Electrostatic Potential in Air-Filled Frame}\label{sec:real}
In this section, we analyze the electrostatic potential in an air-filled frame to investigate how different prior distributions influence GP modeling performance. Electrostatic phenomena are fundamental to many engineering applications, particularly in the design and operation of high-voltage systems, capacitors, and electronic devices. Accurate modeling of electrostatic potential and field distributions is essential for ensuring device reliability and performance. In electrostatics, the electric field $\mathbf{E}$ is related to the scalar potential $V$ by $\mathbf{E}=-\nabla V$. By applying one of Maxwell’s equations, $\nabla \cdot \mathbf{D} = \rho$, and the constitutive relationship $\mathbf{D}=\epsilon \mathbf{E}$, where $\epsilon = \epsilon_r \epsilon_0$ is the absolute dielectric permittivity of the material, $\epsilon_r$ is the relative permittivity of the material, $\epsilon_0$ is the vacuum permittivity which is known as $8.854 \times 10^{-12} \text{F}/\text{m}$, and $\rho$ represents the space charge density, we obtain the Poisson's equation: 
\[
-\nabla \cdot \big(\varepsilon(\mathbf{x})\,\nabla V(\mathbf{x})\big) \;=\; \rho(\mathbf{x})
\quad \text{in } \Omega ,
\]
where $\mathbf{x} \in \Omega \subset \mathbb{R}^2$ denotes the spatial coordinate over the air-filled computational domain $\Omega$. By solving these equations, engineers can predict the behavior of electric fields in devices ranging from capacitors to large-scale high-voltage equipment, where proper insulation and field containment are critical. Refer to \cite{griffiths2023introduction} for further details.

\begin{figure}[!t]
\begin{center}
\includegraphics[width=0.5\textwidth]{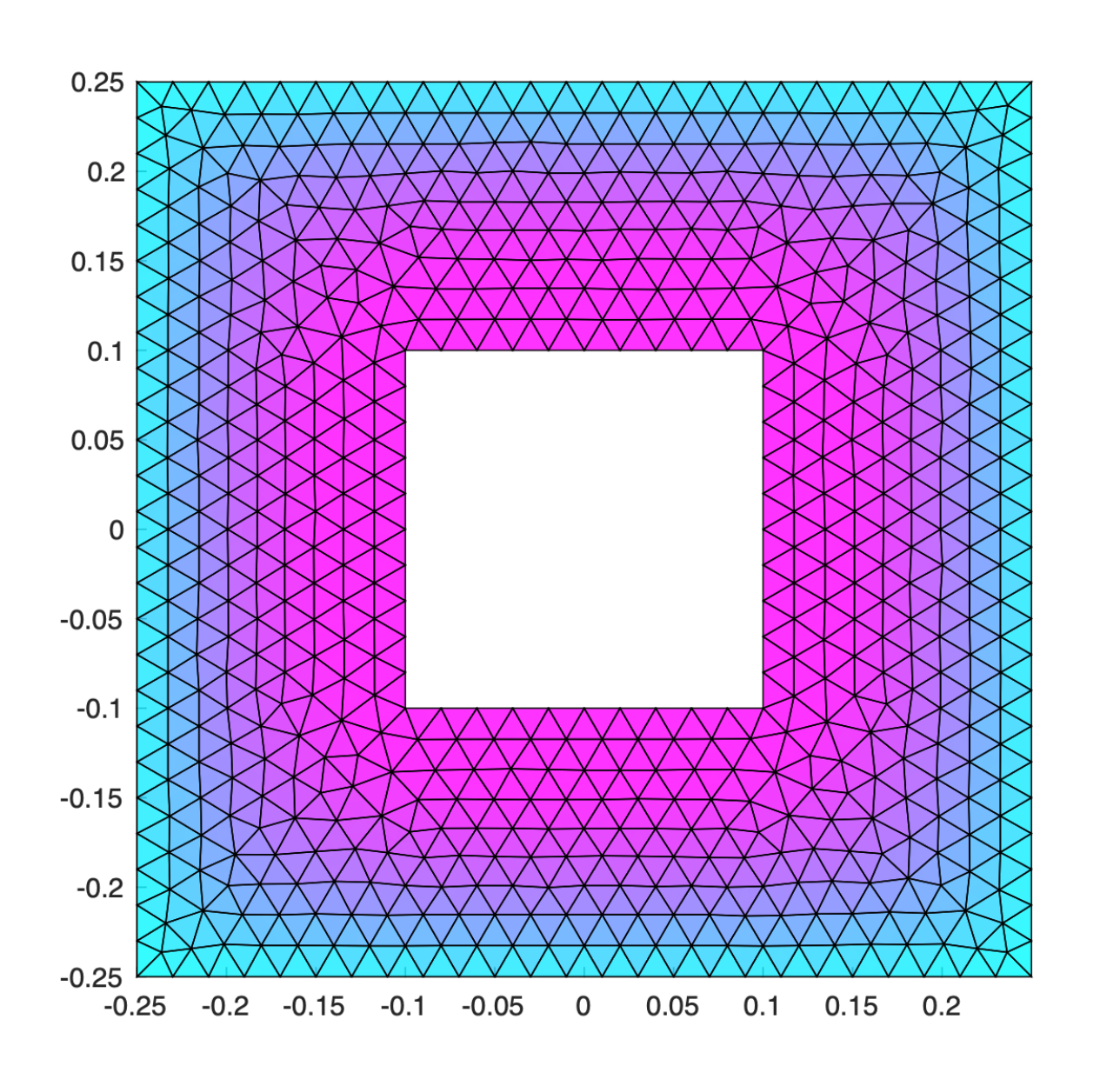} 
\end{center}
\caption{Electrostatic potential distribution in an air-filled annular quadrilateral frame, obtained by finite element simulation with a mesh size of 0.02.}
\label{fig:Electro_Figure}
\end{figure}

Here, we investigate the electrostatic potential distribution in an air-filled annular quadrilateral frame as illustrated in \Cref{fig:Electro_Figure}. The governing Poisson’s equation is solved using the finite element method (FEM) with a uniform mesh of size 0.02, which provides sufficient resolution of the potential field while maintaining computational efficiency. The material properties are characterized by the relative permittivity of air, $\epsilon_r \approx 1.00059$, which is close to unity and has negligible influence on the solution. The boundary conditions are set such that the electrostatic potential is $1000\,\text{V}$ on the inner boundary and $0\,\text{V}$ on the outer boundary. The input variable is the space charge density, $\rho \in [1\times 10^{-6},5\times 10^{-6}] \text{C}/\text{m}^3$, while the outputs of interest are the potential distribution $V(\mathbf{x})$ across all nodes and, in particular, the maximum electric field $\max_{\Omega}\lvert \mathbf{E}(\mathbf{x}) \rvert$, where $\mathbf{E} = -\nabla V$, to assess electrical safety and prevent dielectric breakdown in the air-filled frame.

\begin{figure}[!ht]
    \centering
    \text{RMSE} \\
    \subfloat[Uniform\label{fig:5a}]{\includegraphics[width=0.37\linewidth, height=0.33\linewidth]{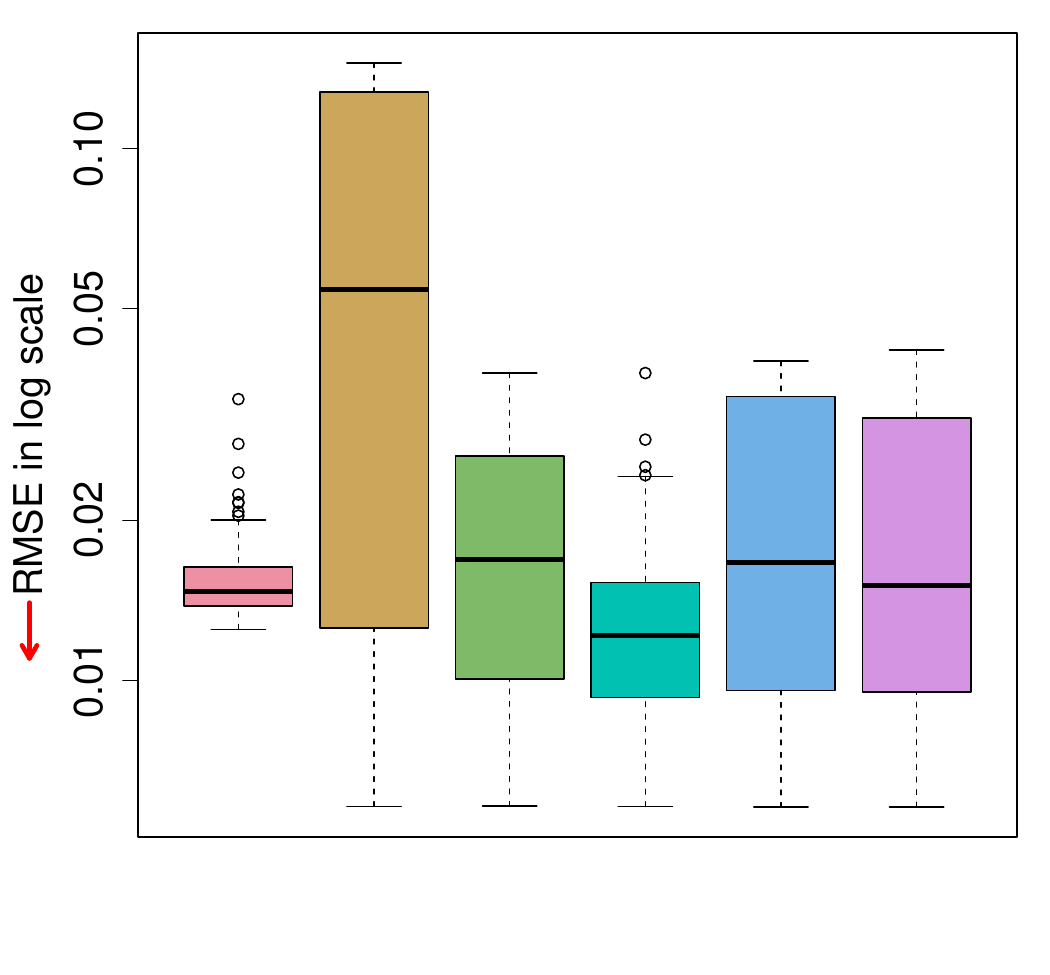}}
     \subfloat[Normal\label{fig:5b}]{\includegraphics[width=0.37\linewidth, height=0.33\linewidth]{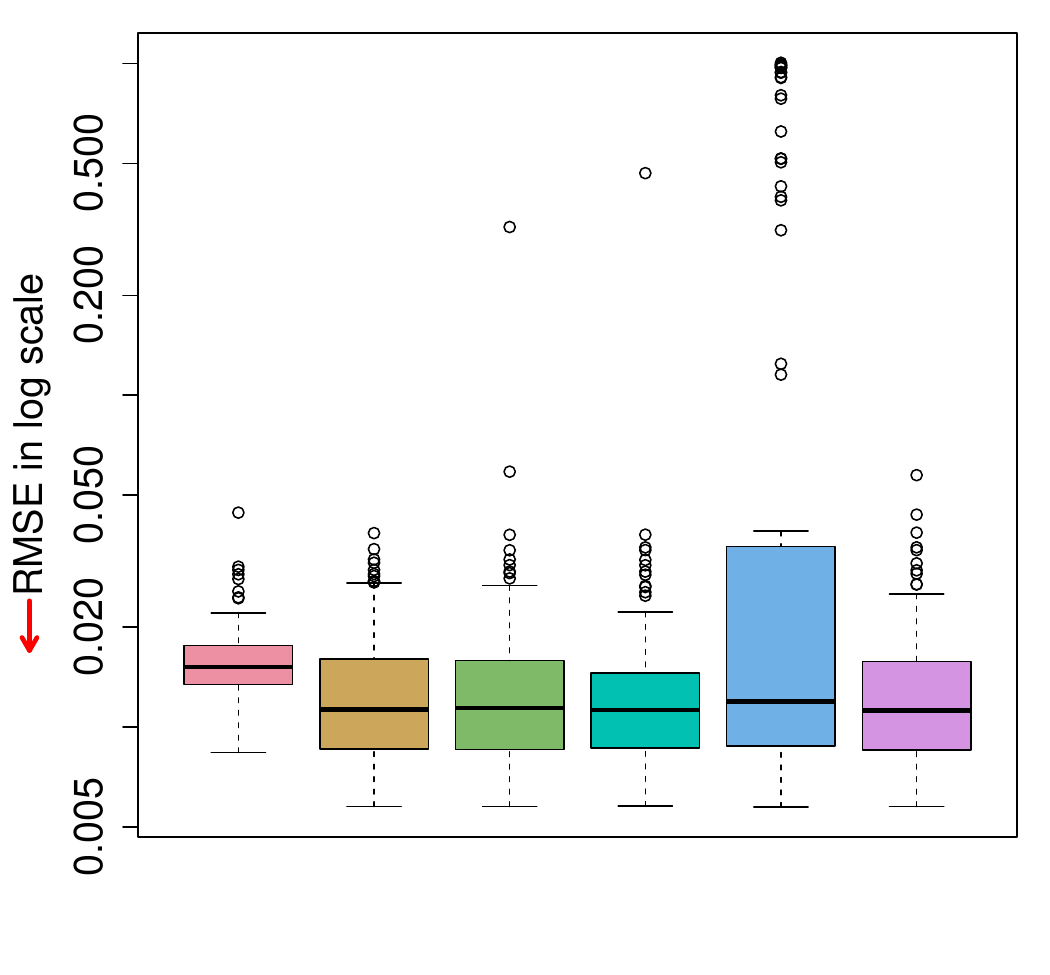}} \\
    \text{CRPS} \\
    \subfloat[Uniform\label{fig:5c}]{\includegraphics[width=0.37\linewidth, height=0.33\linewidth]{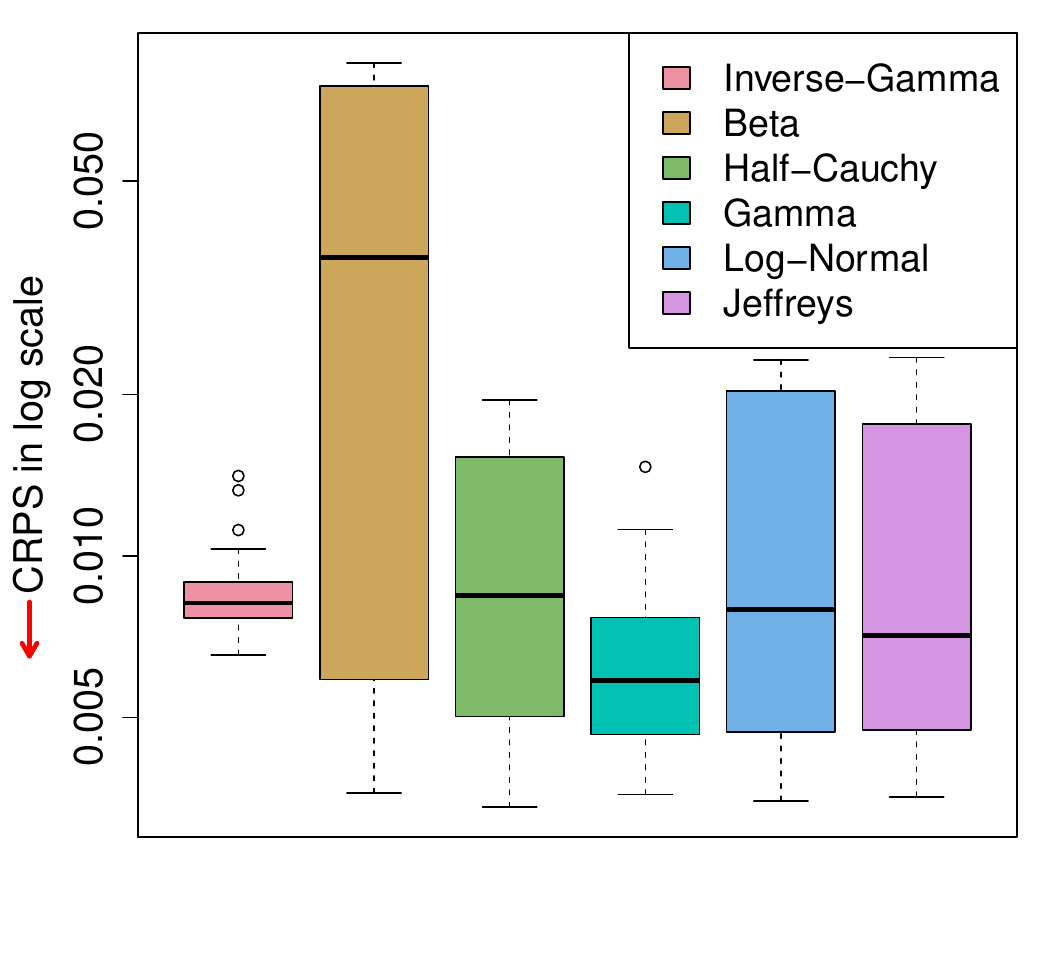}}
 \subfloat[Normal\label{fig:5d}]{\includegraphics[width=0.37\linewidth, height=0.33\linewidth]{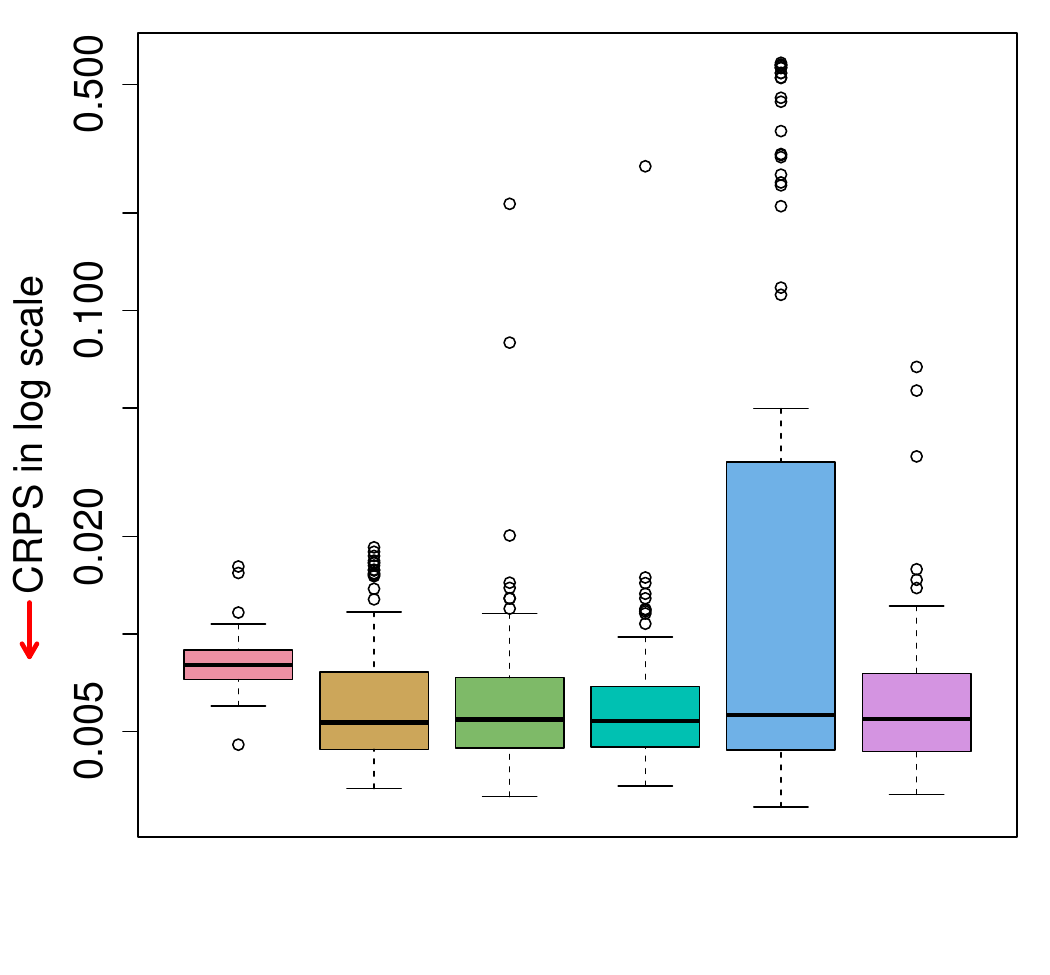}} \\
    \text{PICR} \\
     \subfloat[Uniform\label{fig:5e}]{\includegraphics[width=0.37\linewidth, height=0.33\linewidth]{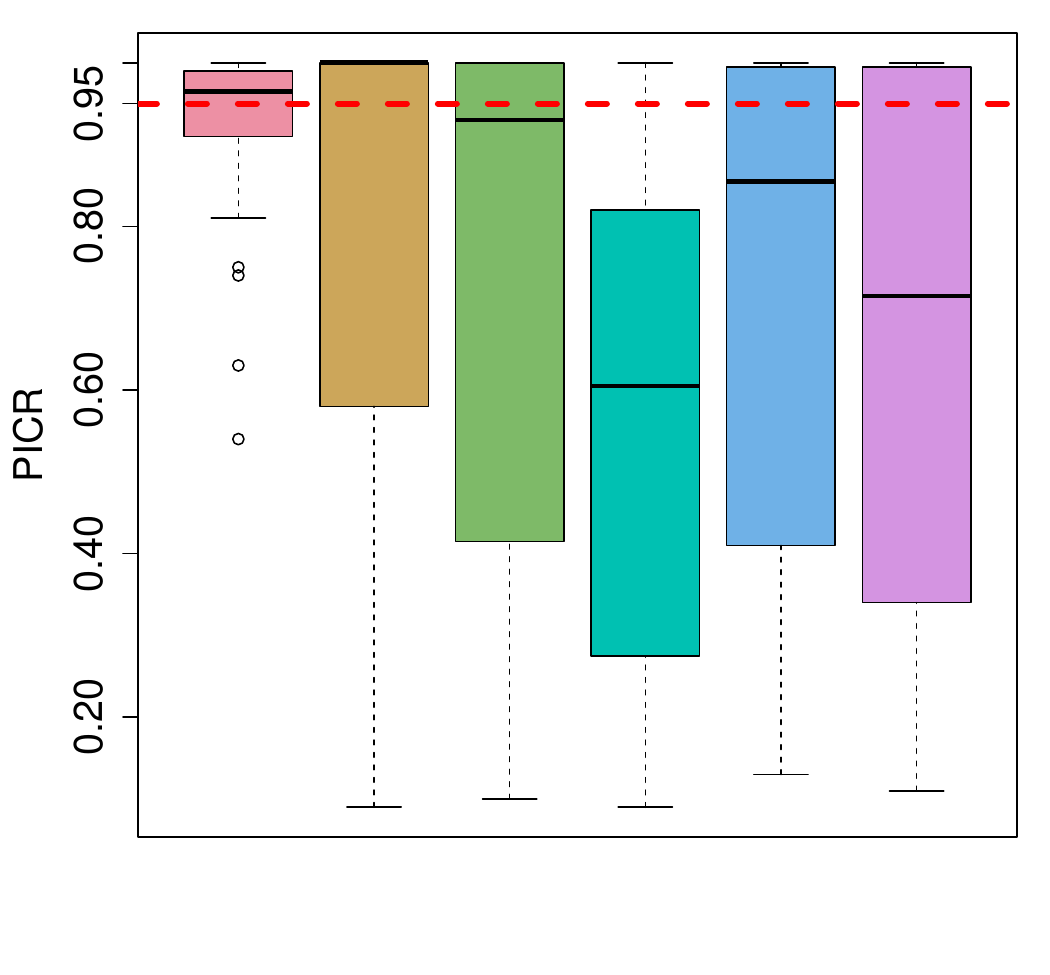}}
     \subfloat[Normal\label{fig:5f}]{\includegraphics[width=0.37\linewidth, height=0.33\linewidth]{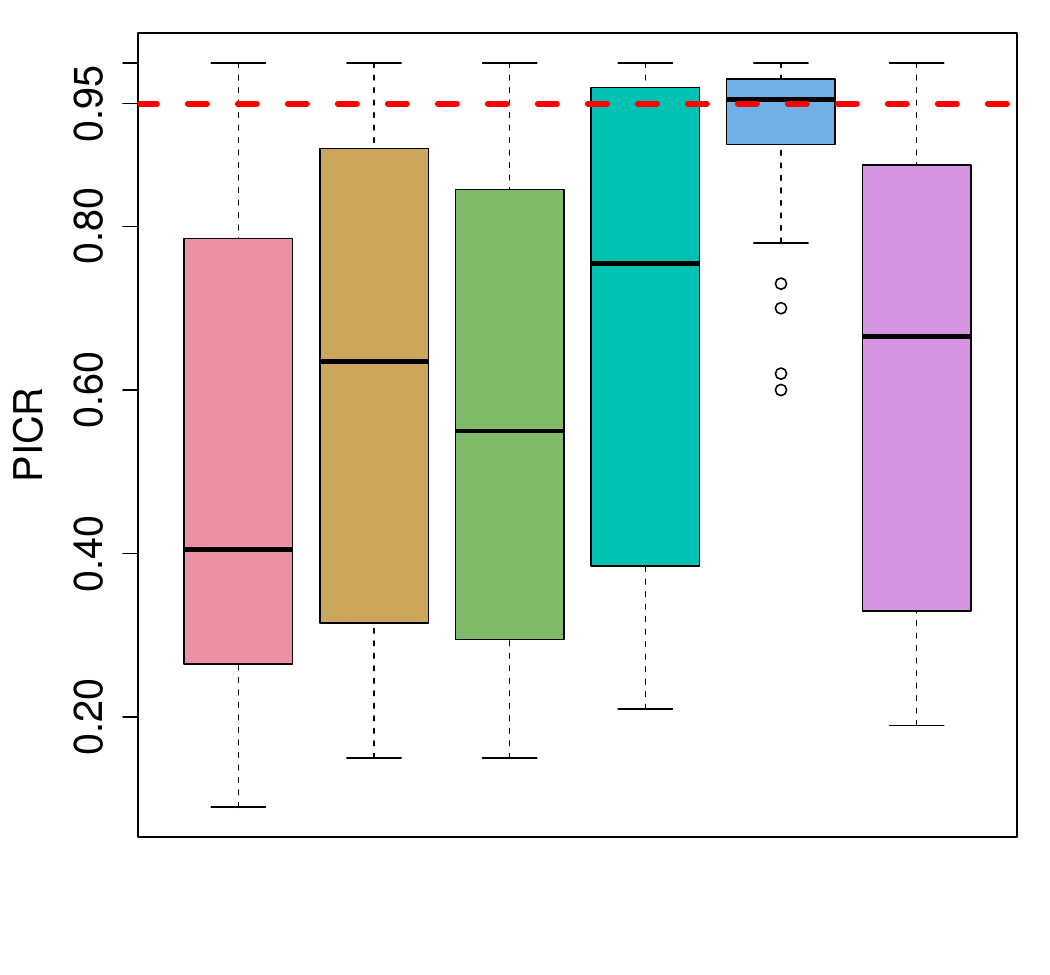}}
    \caption{RMSE (upper), CRPS (middle), and PICR (lower) across 100 repetitions for the electrostatic potential using a uniform proposal distribution (left) and a normal proposal distribution (right).}
    \label{fig:electro}
\end{figure}

The results in \Cref{fig:electro} correspond to the uniform proposal distribution (left) and the normal proposal distribution (right). Under the uniform proposal distribution, the Inverse-Gamma prior yields lower RMSE and CRPS with noticeably smaller variation, and achieves PICR close to 0.95, indicating stable predictive behavior and overall superior performance. In contrast, Beta prior exhibits substantially greater variability and occasional large uncertainties. Although the Gamma prior produces the lowest RMSE and CRPS, it also yields the lowest PICR, implying that the predictive intervals are excessively narrow and suffer from under-coverage. On the other hand, under the normal proposal distribution, all priors except the Log-Normal show relatively comparable RMSE and CRPS values. However, only the Log-Normal prior achieves an average PICR near the nominal 0.95 level, whereas the remaining priors exhibit considerably lower coverage. Notably, the Inverse-Gamma prior yields an average PICR of approximately 0.4, contrast to its behavior under the uniform proposal distribution, where it attains nearly nominal coverage. Consistent with the earlier findings, the results demonstrate that thoughtful specification of both prior and proposal distributions is essential for achieving stable and reliable GP predictions.

\section{Discussion}\label{sec:conclusion}

In this paper, we conducted empirical studies to examine the influence of various prior and proposal distributions for the lengthscale parameter in GPs. The priors considered, which are commonly implemented in various existing software packages, were selected to reflect widely used approaches in practice. The purpose of our analysis was to evaluate these predefined priors and assess their impact on predictive performance. The results from both simulation studies and the real-data analysis demonstrate that predictive performance is highly sensitive to the choice of prior distribution for the lengthscale parameter. Furthermore, we found that careful tuning of the hyperparameters within each prior is essential for achieving robust and reliable predictions. 

While our focus was limited to the lengthscale parameter, other parameters, such as the scale parameter $\tau^2$, also play crucial roles in GPs. Although $\tau^2$ can be obtained in closed form from the likelihood and is often profiled out, prior choice can affect estimation in small-sample or noisy settings. Future work could extend this investigation to include prior specifications for $\tau^2$ to provide a more comprehensive understanding of their effects. It is important to note that our assessment of the influence of prior and proposal distributions on GP performance is not exhaustive. Many alternative priors are available across different software packages and libraries, and further exploration is needed to fully capture the impact of the prior specification within the Bayesian framework for GPs.

\vspace{0.5cm}
\noindent\textbf{Supplemental Materials}
Additional supporting materials are provided in Supplemental Materials, including the R code used to reproduce the results in Sections \ref{sec:studies} and \ref{sec:real}.

\vspace{0.5cm}
\noindent\textbf{Data Availability Statement} The authors confirm that the data supporting the findings of this study are available within its supplementary materials.

%\vspace{0.5cm}
%\noindent\textbf{Acknowledgments} The authors thank Dr. Jonathan Williams of North Carolina State University for providing valuable advice.

\vspace{0.5cm}
\noindent\textbf{Disclosure Statement} No potential competing interests were reported by the authors.
no conflicts of interest to disclose

\bibliography{bib}

\appendix
\section{Fisher information matrix}\label{appendix}

Let $\vect{Y}_n \in\mathbb R^n$ with 
$\vect{Y}_n \sim \mathcal N_n\!\big(\mathbf 0,\; \tau^2 \mathbf{K}\big)$, 
where $\tau^2>0$, and $\boldsymbol{\theta} = (\theta_1,\dots,\theta_d)^\top$. Given the log-likelihood defined in Equation~\ref{eq:loglike}, the score functions are computed as follows:
\begin{comment}
    \paragraph{Log-likelihood.}
\begin{align*}
\ell(\tau^2,\boldsymbol\theta)
&= \log p(Y_n \mid \tau^2,\boldsymbol\theta) \\
&= -\tfrac12\Big\{
\log\lvert \tau^2 \mathbf{K}\rvert + Y_n^\top (\tau^2 \mathbf{K})^{-1} Y_n + n\log(2\pi)
\Big\} \\
&= -\tfrac12\Big\{
n\log \tau^2 + \log|\mathbf{K}| + \tfrac{1}{\tau^2} Y_n^\top \mathbf{K}^{-1} Y_n + n\log(2\pi)
\Big\},
\end{align*}
\end{comment}
\begin{align*}
\frac{\partial \log\mathcal{L}}{\partial \theta_i} = -\frac12\,\mathrm{tr}\left(\mathbf{K}^{-1} \frac{\partial \mathbf{K}}{\partial \theta_i} \right) + \frac{1}{2\tau^2}\, \vect{Y}_n^\top \mathbf{K}^{-1} \frac{\partial \mathbf{K}}{\partial \theta_i} \mathbf{K}^{-1} \vect{Y}_n,
\quad i=1,\dots,d,
\end{align*}
where
\[
\left(\frac{\partial \mathbf{K}}{\partial \theta_i}\right)_{ab}
= \frac{\partial \mathbf{K}_{ab}}{\partial \theta_i}
= \mathbf{K}_{ab}\,\frac{(x_{a,i}-x_{b,i})^2}{\theta_i^2}.
\]

%\paragraph{Fisher information.}
Let $\mathcal{I}(\boldsymbol{\theta})$ be the Fisher information matrix for $\boldsymbol{\theta}$.
Taking expectations under $\vect{Y}_n \sim\mathcal N(0,\tau^2 \mathbf{K})$ yields 
\begin{align*}
\mathcal I_{\tau^2,\tau^2}
&= \frac{n}{2} \frac{1}{(\tau^2)^2},\\
\mathcal I_{\tau^2,\boldsymbol\theta_i}
&= \frac{1}{2\tau^2}t_i,\\
\mathcal I_{\boldsymbol\theta_i,\boldsymbol\theta_j}
&= \frac12 S_{ij},\quad i,j=1,\dots,d,
\end{align*}
where $t_i=\mathrm{tr}\left(\mathbf{K}^{-1} \frac{\partial \mathbf{K}}{\partial \theta_i}\right)$ and $\mathbf{S}=(S_{ij})_{i,j=1}^d=(\mathrm{tr}\left(\mathbf{K}^{-1} \frac{\partial \mathbf{K}}{\partial \theta_i} \mathbf{K}^{-1} \frac{\partial \mathbf{K}}{\partial \theta_j}\right))_{i,j=1}^d$.

Using the Fisher information, the joint Jeffreys prior for $(\tau^2,\boldsymbol{\theta})$ is proportional to
\begin{align*}
    p(\tau^2, \boldsymbol \theta) &\propto \frac{1}{\tau^2} \left| \mathbf{S}-\frac{1}{n} \mathbf{t} \mathbf{t}^\top \right|^{\frac{1}{2}}.
\end{align*}

Thus, the marginal Jeffreys prior for $\boldsymbol{\theta}$ is
\begin{align*}
    p(\boldsymbol \theta) \propto \left| \mathbf{S}-\frac{1}{n} \mathbf{t} \mathbf{t}^\top \right|^{\frac{1}{2}}.
\end{align*}

\iffalse
\newpage
\setcounter{page}{1}
\bigskip
\bigskip
\bigskip
\begin{center}
{\Large\bf Supplementary Materials for ``Influence of Prior Distributions on Gaussian Process Hyperparameter Inference''}\\
\if1\blind\vspace{5mm}
  \centering{Ayumi Mutoh \vspace{0.1in}\\
        Department of Statistics, North Carolina State University\\
  and\\
  Junoh Heo\footnote{Corresponding author. \href{mailto:heojunoh@msu.edu}{heojunoh@msu.edu}} \vspace{0.1in}\\
  Department of Statistics and Probability, Michigan State University}\fi
\end{center}

\medskip

\setcounter{section}{0}
\setcounter{equation}{0}
\setcounter{figure}{0}
\def\theequation{S\arabic{section}.\arabic{equation}}
\def\thesection{S\arabic{section}}
\def\thefigure{S\arabic{figure}}
\def\thetable{S\arabic{table}}
\fi

\end{document}